\documentclass[twocolumn,showpacs,preprintnumbers,amsmath,amssymb]{revtex4}

\usepackage{graphicx}
\usepackage{dcolumn}
\usepackage{bm}

\begin{document}

\title{A Universal Map for Fractal Structures in Weak Solitary Wave Interactions}

\author{Yi Zhu$^{1}$, Richard Haberman$^{2}$, Jianke Yang$^{3}$}
\affiliation{%
$^{1}$Zhou Pei-Yuan Center for Applied Mathematics,
Tsinghua University, Beijing 100084, China \\
$^{2}$Department of Mathematics, Southern Methodist
University, Dallas, Texas 75275, USA\\
$^{3}$Department of Mathematics and Statistics, University of
Vermont, Burlington, VT 05401, USA
}%

\date{Received: ***}

\begin{abstract}
Fractal scatterings in weak solitary wave interactions is analyzed
for generalized nonlinear Schr\"odiger equations (GNLS). Using
asymptotic methods, these weak interactions are reduced to a
universal second-order map. This map gives the same fractal
scattering patterns as those in the GNLS equations both
qualitatively and quantitatively. Scaling laws of these fractals are
also derived.
\end{abstract}

\pacs{42.65.Tg, 05.45.Yv, 42.81.Dp}
\maketitle

Solitary wave interactions is a fascinating mathematical phenomenon,
and it arises in numerous physical applications such as water waves
and nonlinear optics \cite{Ablowitz,Hasegawa,Kivshar}. Strong
interactions occur when two solitary waves are initially far apart
but move toward each other at moderate or large speeds. Weak
interactions would occur if the two waves are initially well
separated, and their relative velocities are small. In integrable
wave equations, strong interactions of solitary waves are elastic,
and their weak interactions exhibit interesting but simple dynamics
\cite{Ablowitz,Hasegawa,Karpman_Solovev}. In non-integrable systems,
however, solitary wave interactions can be extremely complicated.
Indeed, one of the most important developments in the nonlinear wave
theory in recent years is the discovery of fractal scatterings of
solitary wave interactions in non-integrable equations
\cite{Campbell,anninos,Kivshar_Fei,YangTan,Dmitriev,zhuyang,Goodman_Haberman}.
On the analysis of fractal scatterings, some progress has been made.
For strong interactions, various collective-coordinate ODE models
based on qualitative variational methods have been derived and
analyzed \cite{anninos,Kivshar_Fei,YangTan,Goodman_Haberman}. From
the variational ODEs, a separatrix map was derived, showing chaotic
scatterings [11]. On weak interactions, a simple
asymptotically-accurate ODE model was derived for the generalized
NLS equations \cite{zhuyang}. This ODE system offered the first
glimpse of universal fractal scatterings in weak wave interactions,
but these fractal patterns were not analyzed.

In this letter, we analyze fractal scattering patterns in weak
solitary wave interactions for generalized nonlinear Schr\"odiger
equations. Using asymptotic methods, we reduce these weak
interactions to a simple second-order map which contains no free
parameters. It is shown that this universal map gives a complete
characterization of fractal structures in these wave interactions.
In addition, the scaling laws of these fractals for different
initial conditions are analytically derived. These results provide a
deep understanding of weak solitary-wave interactions for various
physical applications.

The generalized nonlinear Schr\"odinger equations we consider in
this paper are
\begin{equation} \label{GNLS}
iU_t+U_{xx}+F(|U|^2)U=0,
\end{equation}
where $F(\cdot)$ is a general function. These equations govern
various physical wave phenomena in nonlinear optics, fiber
communications and fluid dynamics \cite{Ablowitz,Hasegawa,Kivshar}.
This equation admits solitary waves of the form
$U=\Phi(x-\xi)e^{i\phi}$, where $\Phi(\theta)$ is a localized
positive function, $\xi=Vt+x_0$ is the wave's center position, and
$\phi=\frac{1}{2}V(x-\xi)+(\beta+\frac{1}{4}V^2)t-\eta_0$ is the
wave's phase. This wave has four free parameters: velocity $V$,
amplitude parameter $\beta$, initial position $x_0$, and initial
phase $\eta_0$. In weak interactions, two such solitary waves are
initially well separated and have small relative velocities and
amplitude differences. Then they would interfere with each other
through tail overlapping. When time goes to infinity, they either
separate from each other or form a bound state. The exit velocity,
defined as $\Delta V_\infty=|V_2-V_1|_{t\to \infty}$, depends on the
initial conditions of the two waves.

\begin{figure}[t]
\includegraphics[width=80mm,height=50mm]{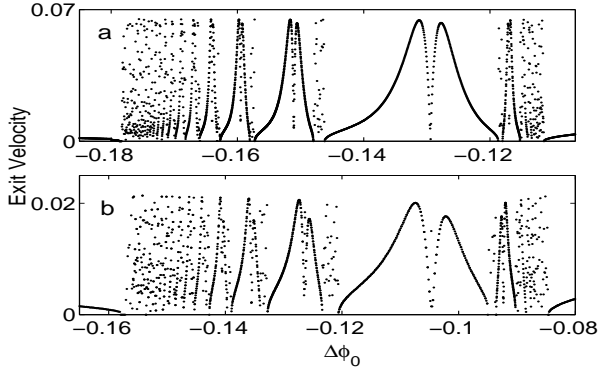}
\caption{Exit velocity versus initial phase difference graphs for
initial conditions (\ref{ic}): (a) cubic-quintic nonlinearity; (b)
quadratic-cubic nonlinearity.\label{fig1}}
\end{figure}

To study weak interactions in Eq. (\ref{GNLS}), we select two
different nonlinearities which are cubic-quintic and quadratic-cubic
respectively:
\begin{equation}\label{cqNL}
F(|U|^2)=|U|^2+\gamma |U|^4,  \quad  F(|U|^2)=|U|^2+\delta |U|.
\end{equation}
Here $\gamma$ and $\delta$ are real parameters. To illustrate
results of weak interactions, we take $\gamma=0.0003$ and
$\delta=-0.0015$. The initial conditions are taken as
\begin{equation} \label{ic}
x_{0,1}=-x_{0,2}=-5, \; V_{0,k}=0, \; \beta_{0,k}=1 \: (k=1, 2),
\end{equation}
$\phi_{0,1}=0$, and the initial phase difference $\Delta
\phi_0=\phi_{0,2}-\phi_{0,1}$ is used as the control parameter. In
our numerical simulations of Eq. (\ref{GNLS}), the discrete Fourier
transform is used to evaluate the spatial derivative
$\partial_{xx}$, while the fourth-order Runge-Kutta method is used
to advance in time. The exit velocity $\Delta V_\infty$ versus
$\Delta \phi_0$ graphs for these two nonlinearities are plotted in
Fig. \ref{fig1}. These graphs are fractals \cite{zhuyang}. It is
amazing that these fractals appear for such small values of $\gamma$
and $\delta$, where Eq. (\ref{GNLS}) is simply a weakly perturbed
NLS equation. Notice that the fractals for these two different
nonlinearities are very similar, signaling their universality in
weak wave interactions.

To analyze this fractal-scattering phenomena, the Karpman-Solov'ev
method \cite{Karpman_Solovev} was applied, and the following simple
set of dynamical equations for solitary wave parameters were derived
\cite{zhuyang}:
\begin{equation}
\zeta_{\tau\tau}=\cos{\psi}e^{\zeta},\quad
\psi_{\tau\tau}=(1+\varepsilon)\sin{\psi}e^{\zeta}.
 \label{eqDyfinal}
\end{equation}
Here $\psi=\Delta\phi$, $\zeta=-\sqrt{\beta}\Delta\xi$,
\begin{eqnarray} \label{scaling1}
\tau=\sqrt{16\beta^{3/2} c^2/P}\; t, \;\; \varepsilon=P/(2\beta
P_\beta)-1,
\end{eqnarray}
$\Delta \xi$ and $\Delta \phi$ are the distance and phase difference
between the two waves, $\beta=(\beta_{1,0}+\beta_{2,0})/2$, $c$ is
the tail coefficient of the solitary wave with propagation constant
$\beta$, and $P(\beta)$ is the power function of the wave. These
reduced ODEs capture fractal scatterings of weak wave interactions
such as in Fig. 1 both qualitatively and quantitatively
\cite{zhuyang}, and they represent an important first step toward
the understanding of these phenomena. However, the fractal
structures in the PDEs (\ref{GNLS}) and ODEs (\ref{eqDyfinal}) have
not been analyzed previously. Below we give a complete
characterization for the first time of this fractal scattering by
analyzing the ODEs (\ref{eqDyfinal}).

If $\varepsilon=0$, Eq. (\ref{eqDyfinal}) is integrable. It has two
conserved quantities, energy $E$ and momentum $M$:
\begin{equation}\label{EM}
E=(\dot{\zeta}^2-\dot{\psi}^2)/2-e^\zeta\cos{\psi}, \quad M =
\dot{\zeta}\dot{\psi}-e^\zeta\sin{\psi}.
\end{equation}
Introducing two complex quantities
\begin{equation}  \label{Cdef}
C =\sqrt{(E+iM)/2}, \; F=
-\mbox{acoth}[(\dot{\zeta}+i\dot{\psi})/2C)]/C,
\end{equation}
the analytical solution of Eq. (\ref{eqDyfinal}) can be found to be
\begin{equation}
Y(\tau)=\ln\left[2C^2_{0}\textrm{csch}^2
C_{0}(\tau-\tau_0+F_0)\right],\label{GSolu1}
\end{equation}
where $Y=\zeta+i\psi$, and $C_0$, $F_0$ are the initial values of
$C$ and $F$. The third conserved quantity of Eq. (\ref{eqDyfinal})
is $\mbox{Im}(F)$. Behaviors of the above integrable solutions
should be noted. When $E>0$, or $E\leq 0$ but $M\neq 0$, $\zeta\to
-\infty$ (a degenerate saddle point) as $\tau\to \infty$, and thus
these solutions are escape orbits. When $E<0$ and $M=0$, the orbits
are periodic with period $T_p=\sqrt{2}\pi / \sqrt{|E|}$. Orbits with
$E=M=0$ separate the escape orbits from the periodic ones, hence we
call them separatrix orbits. The formulae for separatrix orbits can
be readily found to be
\begin{eqnarray}\label{sep2}
Y_s(\tau)=-\ln[\sigma A ^{-1/2}i+(\tau-\tau_M)/\sqrt{2}]^2.
\end{eqnarray}
Here $A=e^{\zeta_M}$, $\zeta_M$ is the maximum of $\zeta(\tau)$,
$\tau_M$ is the time when $\zeta=\zeta_M$, and $\sigma$ is the sign
of $\dot{\psi}$ at $\tau=\tau_M$.

In the general case where $\varepsilon \ne 0$, Eq. (\ref{eqDyfinal})
is still a Hamiltonian system with the conserved Hamiltonian
\begin{equation}\label{Heps}
H(\zeta,\dot{\zeta},\psi,\dot{\psi})=E+\varepsilon
\dot{\psi}^2/[2(1+\varepsilon)],
\end{equation}
where $E$ is given in Eq. (\ref{EM}). But $E, M$ and $\mbox{Im}(F)$
are not conserved anymore. In order to determine the fractal
structures as shown in Fig. 1, we need to calculate the exit
velocity, which corresponds to $|\dot{\zeta}_\infty|$ for Eq.
(\ref{eqDyfinal}). If the orbit has non-zero exit velocity, then
from Eqs. (\ref{EM}) and (\ref{Heps}), we find that
\begin{eqnarray} \label{exit_V}
|\dot{\zeta}_\infty|=\sqrt{H+\sqrt{H^2+M^2_\infty/(1+\varepsilon)}}.
\end{eqnarray}
Since $H$ is conserved, to get $|\dot{\zeta}_\infty|$, we only need
to find $M_\infty$. For arbitrary values of $\varepsilon$, it is
impossible to calculate $M_\infty$ analytically. However, when
$\varepsilon \ll 1$ as in Fig. 1 (where $\varepsilon=0.001$ for both
nonlinearities), the calculation of $M_\infty$ can be done. In this
case, Eq. (\ref{eqDyfinal}) is weakly perturbed from the integrable
case ($\varepsilon=0$), thus we will use asymptotic techniques in
our calculations below.

\begin{figure}[t]
\includegraphics[width=40mm,height=30mm]{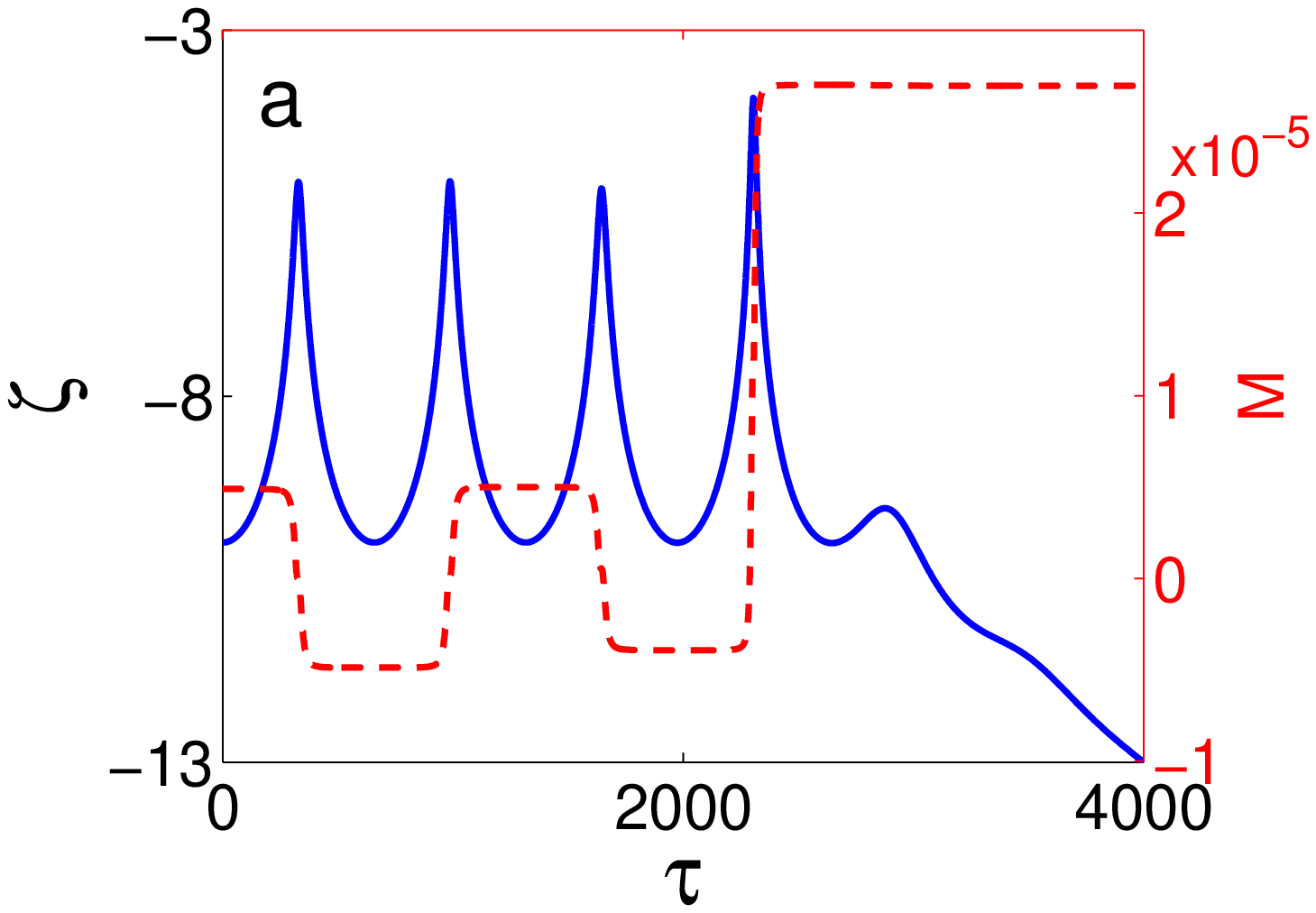}
\includegraphics[width=40mm,height=30mm]{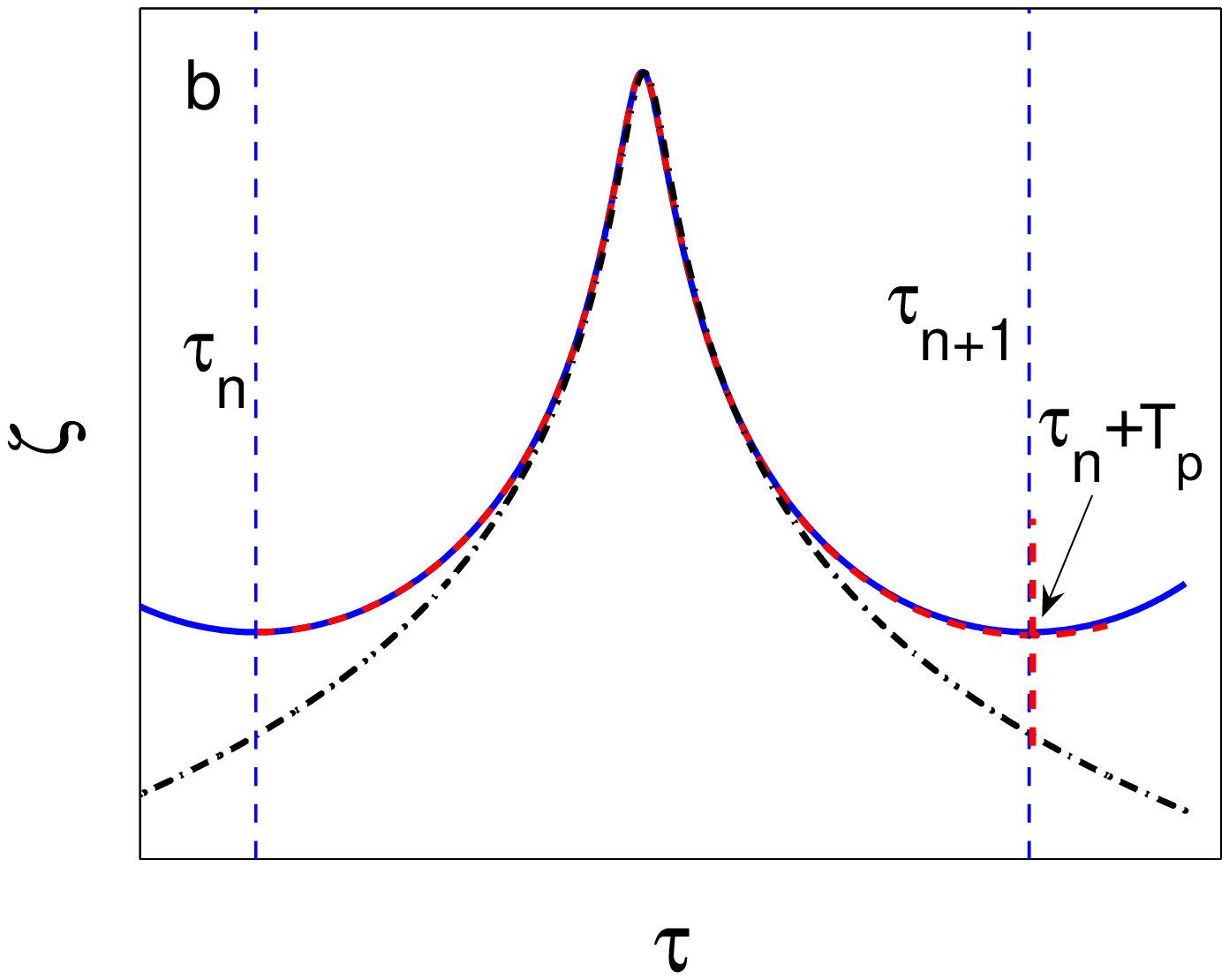}
\caption{(color online) (a) A typical $\zeta$ solution (blue solid)
when $\Delta \phi_0$ is in the fractal region, and the corresponding
$M(\tau)$ curve (red dashed); (b) plots of the perturbed solution
$\zeta(\tau)$ (blue solid), the unperturbed solution (red dashed),
as well as the separatrix solution (black dash-dotted) from the
$n$-th saddle approach to the next.}
\end{figure}

To motivate our analysis, we first illustrate in Fig. 2(a) a typical
$\zeta$ solution when the initial condition lies in the sensitive
region of Fig. 1. We see that $\zeta$ undergoes several large
oscillations, then escapes to $-\infty$. Each oscillation
corresponds to a ``bounce" in the two-wave interactions. These
bouncing sequences are the key to the existence of fractal
structures, similar to other physical systems
\cite{Campbell,anninos,Kivshar_Fei,YangTan,Dmitriev}. Each local
minimum of $\zeta$ will be called a saddle approach
\cite{Goodman_Haberman}. The corresponding $M(\tau)$ curve is also
plotted in Fig. 2(a). We see that $M$ changes very little near a
saddle approach, but changes significantly near maxima of $\zeta$.
Below we will calculate the change in $M$ from one saddle approach
to another. It turns out the $M$ formula will be coupled to $E$ and
$F$, thus we need to calculate the changes in $E, M$ and $F$
simultaneously. To carry out these calculations, we notice two
facts. One is that from one saddle approach to the next, the
perturbed and unperturbed (i.e. integrable) solutions remain close
to each other since $\varepsilon \ll 1$. The other fact is that at
each saddle approach, $E_n,M_n\ll 1$. This is so since initial
$E_0,M_0$ are always small for weak wave interactions, and they will
remain small when $\varepsilon\ll 1$. To simplify our analysis, we
also assume that at each saddle approach, $M_n/E_n\ll 1$. This
assumption is satisfied for many initial conditions such as
(\ref{ic}).

Now we calculate $E, M$ and $F$ from the $n$-th saddle approach to
the next. From Eqs. (\ref{eqDyfinal}) and (\ref{EM}), we get
\begin{eqnarray}\label{DMformal}
\Delta
M_n=\int^{\tau_{n+1}}_{\tau_n}\frac{dM}{d\tau}d\tau=\int^{\tau_{n+1}}_{\tau_n}\varepsilon
e^\zeta\sin{\psi}\dot{\zeta}d\tau.
\end{eqnarray}
In view of the first fact above, the perturbed orbit in the above
formula can be approximated by the integrable orbit [see Fig. 2(b)].
Due to the second fact, we can further approximate the integrable
orbit by a separatrix orbit. Notice that the separatrix orbit
(\ref{sep2}) has three parameters. To select the appropriate
parameters in the separatrix, we note that most contributions to the
integral of (\ref{DMformal}) come from the $\zeta$-maximum region,
thus it is natural to ask the separatrix solution to have the same
$\zeta$-maximum point as the integrable solution [see Fig. 2(b)].
Then in view of Eq. (\ref{GSolu1}), the above requirement selects
$\sigma$ and $A$ in the separatrix (\ref{sep2}) as
\begin{eqnarray}
&&\sigma_n=\mbox{sgn}\left\{\mbox{Im}\left[-2C_{n}\textrm{\mbox{coth}}(C_{n}(\tau^*_u+F_n))\right]\right\};\label{sigma}\\
&&A_n=2\left|C^2_{n}\textrm{csch}^2\left[C_{n}(\tau^*_u+F_n)\right]\right|.
\label{ezetam}
\end{eqnarray}
Here $\tau_n+\tau^*_u$ is the time the unperturbed solution
$\zeta_u$ reaches the maximum. Due to the assumption $M_n/E_n\ll 1$,
the unperturbed solution (to leading order) is a periodic solution
with period $T_p=\sqrt{2}\pi/\sqrt{|E_n|}$. Thus $\tau^*_u = T_p/2$,
and $\tau_{n+1}=\tau_n+T_p$. Utilizing the above results and
noticing $T_p\gg 1$, Eq. (\ref{DMformal}) is asymptotically
approximated by
\begin{eqnarray}
&&\Delta M_n= \varepsilon\int^{+\infty}_{-\infty}
e^{\zeta_s}\sin{\psi_s}\dot{\zeta}_sd\tau =\sigma_n\varepsilon \pi
A_n/2.  \label{DM}
\end{eqnarray}
By similar calculations and utilizing the symmetry properties of the
separatrix solution (\ref{sep2}), we find that
\begin{eqnarray}
\Delta E_n=\varepsilon\int^{+\infty}_{-\infty}
e^{\zeta_s}\sin{\psi_s}\dot{\psi}_sd\tau=0. \label{DE}
\end{eqnarray}

To calculate $F_{n+1}$, notice from Eqs. (\ref{eqDyfinal}) and
(\ref{Cdef}) that $F$ satisfies a linear inhomogeneous ODE:
\begin{equation}
\frac{dF}{d\tau}=-\frac{\dot{E}+i\dot{M}}{2(E+i M)}F+D, \label{fdot}
\end{equation}
where $D$ is a function of $(\zeta, \psi)$ whose expression is easy
to obtain. The homogeneous solution of this ODE is $C^{-1}(\tau)$.
Thus by using the method of variation of parameters, we can
integrate the inhomogeneous ODE (\ref{fdot}) from $\tau_n$ to
$\tau_{n+1}$ and get
\begin{eqnarray}
F_{n+1}=F_n\frac{C_{n}}{C_{n+1}}+\frac{\int^{\tau_{n+1}}_{\tau_n}\sqrt{E+iM}
Dd\tau}{\sqrt{E_{n+1}+iM_{n+1}}}.
\end{eqnarray}
Due to the second fact of $E_n,M_n\ll 1$, we can approximate the
solution $(\zeta, \psi)$ in $D$ by the separatrix solution
(\ref{sep2}). Then to leading order in $\varepsilon$, we get
\begin{eqnarray} \label{Fn1}
F_{n+1}=F_n\frac{C_{n}}{C_{n+1}}+\frac{\left.(\hat{\tau}+i\alpha)
\sqrt{E+iM}\right|^{\tau_{n+1}}_{\tau_n}}{\sqrt{E_{n+1}+iM_{n+1}}}.
\end{eqnarray}
Here $\hat{\tau}=\tau-\tau_n-T_p/2$,  and $\alpha=\sqrt{2}\sigma_n
A_n^{-1/2}$.

Iteration equation (\ref{Fn1}) is quite complicated. Below, we
simplify it. From Eq. (\ref{DE}), we get $E_n=E_0$. Under our
assumption of $M_n/E_n\ll 1$, to leading order, Eq. (\ref{Fn1})
becomes
\begin{eqnarray}
F_{n+1}=\frac{\pi }{\sqrt{2|E_0|}}+(F_n+\frac{\pi
}{\sqrt{2|E_0|}})(1-\frac{M_{n+1}-M_n}{2E_0}i). \label{Fn+1}
\end{eqnarray}
At the initial saddle approach, we find from Eq. (\ref{Cdef}) that
$\mbox{Re}(F_0)=-\pi/\sqrt{2|E_0|}$. Then solving Eq. (\ref{Fn+1}),
we get
\begin{eqnarray}
F_n=\pi(2|E_0|)^{-1/2}\left[2n-1-i\hspace{0.02cm} S_{n}M_n/2E_0
\right], \label{Fn2}
\end{eqnarray}
where $S_{n+1}M_{n+1}=2nM_{n+1}-(2n-S_n)M_n$. Now we introduce a new
variable $Q_n$:
\begin{eqnarray}
Q_n - 2n M_n  =  -S_nM_n =  - (2|E_0|)^{3/2} \mbox{Im}[F_n]/\pi.
\end{eqnarray}
Then substituting formula (\ref{Fn2}) into (\ref{sigma}),
(\ref{ezetam}), keeping only their leading order terms in $M_n/E_n$,
and putting the resulting expressions into (\ref{DM}), we obtain the
simplified iteration equations as
\begin{eqnarray}
&&M_{n+1}=M_n-\mbox{sgn}(Q_n)8|E_0|^3\varepsilon/\pi Q_n^2,  \label{map1}\\
&&Q_{n+1}=Q_n+2M_{n+1}. \label{map2}
\end{eqnarray}
These equations are derived asymptotically near the separatrix orbit
(\ref{sep2}), and will be called the separatrix map. This map can be
further normalized. Let
\begin{equation} \label{scaling2}
G=8|E_0|^3\varepsilon/\pi, \; \tilde{M}_n=G^{-1/3}M_n, \;
\tilde{Q}_n=G^{-1/3}Q_n,
\end{equation}
then the normalized separatrix map is
\begin{eqnarray}
\tilde{M}_{n+1}=\tilde{M}_n-\frac{\mbox{sgn}(\varepsilon\tilde{Q}_n)}{\tilde{Q}_n^2},
\label{nmap1}
\\ \tilde{Q}_{n+1}=\tilde{Q}_n+2\tilde{M}_{n+1}.   \hspace{0.5cm} \label{nmap2}
\end{eqnarray}
This is a simple but important second-order area-preserving map, and
it does not have any parameters in it (except a sign of
$\varepsilon$). This \emph{universal} map governs weak two-wave
interactions in generalized NLS equations (\ref{GNLS}).


\begin{figure}[t]
\includegraphics[width=40mm,height=30mm]{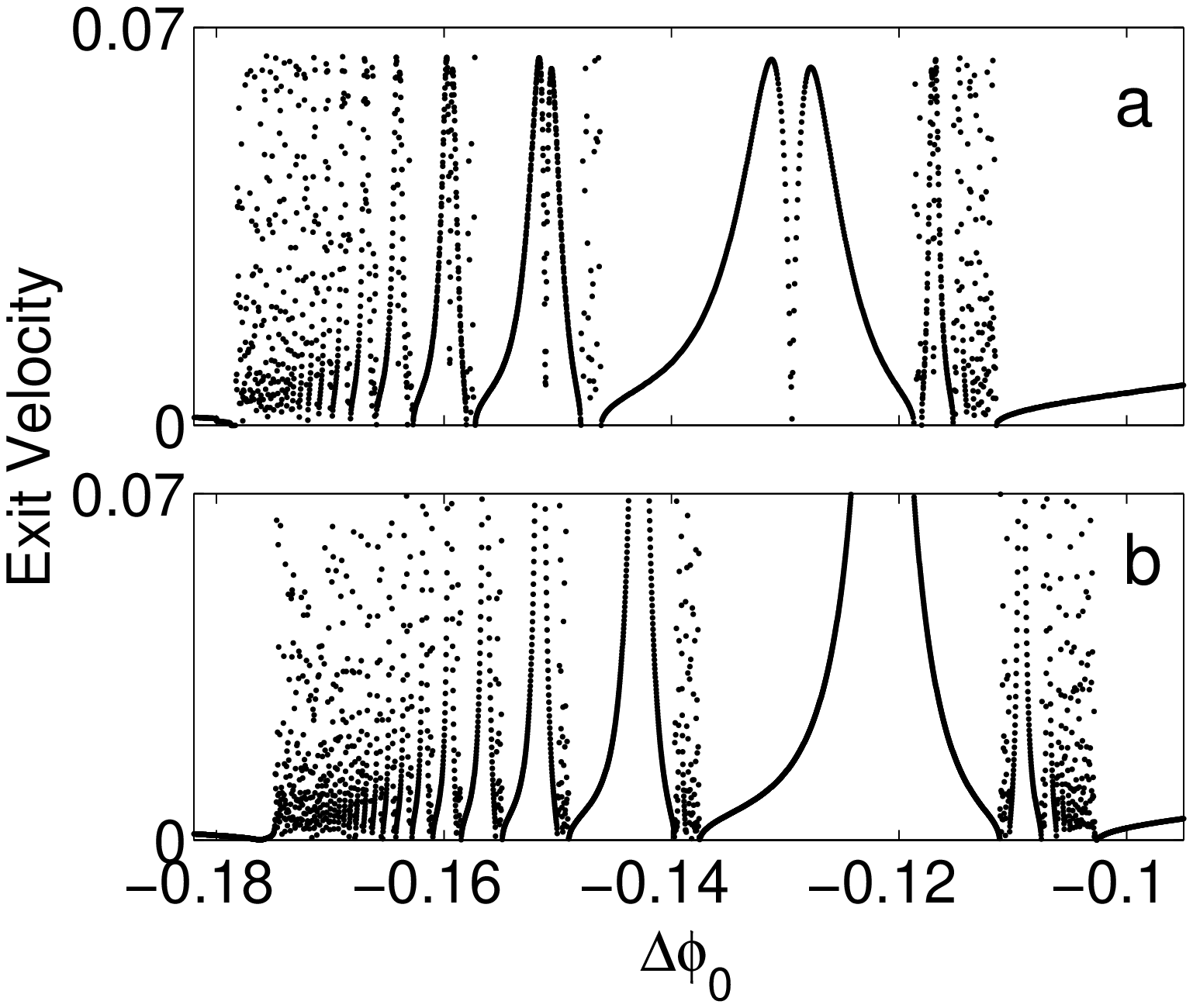}
\includegraphics[width=40mm,height=30mm]{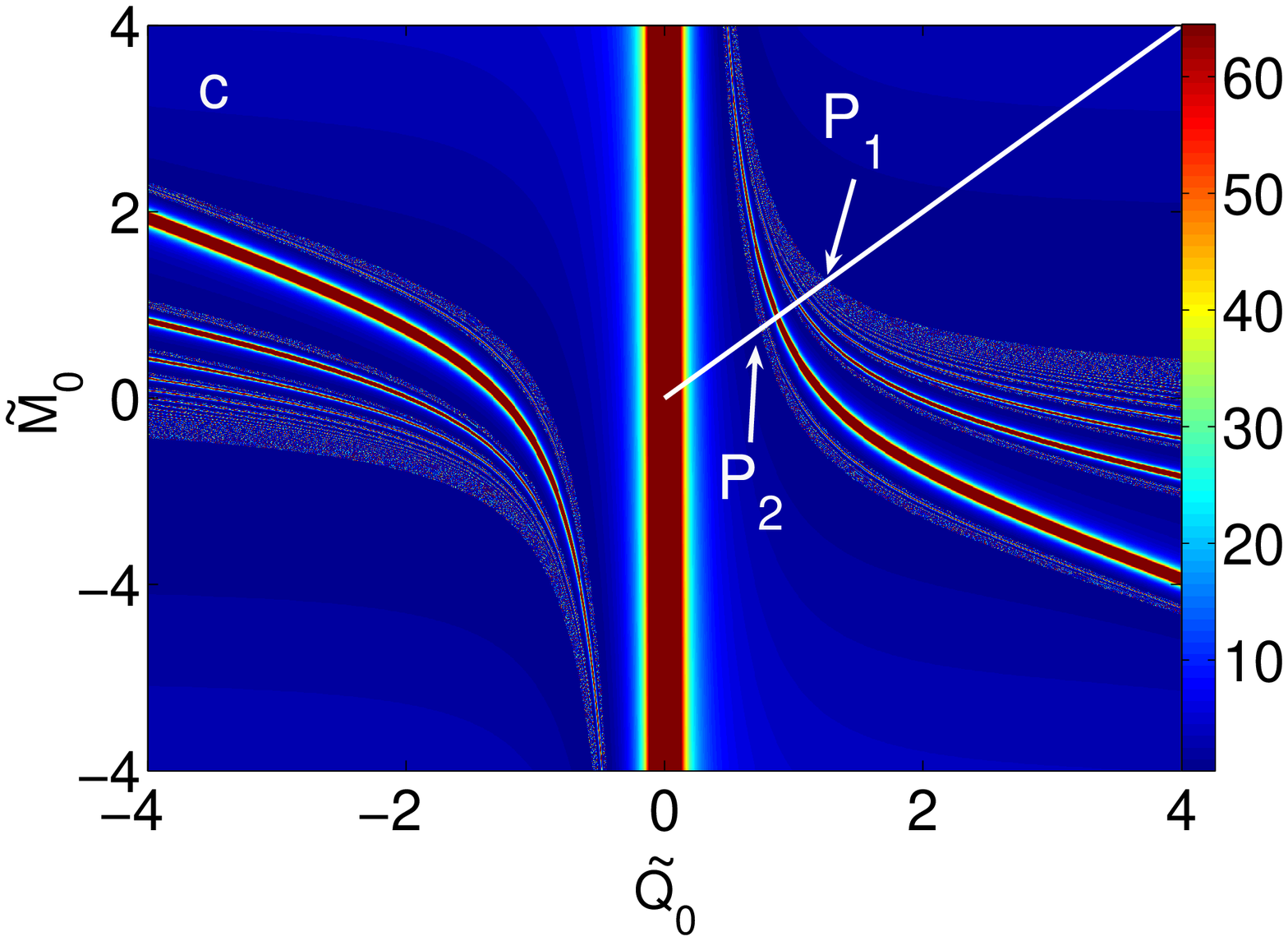}
 \includegraphics[width=40mm,height=31mm]{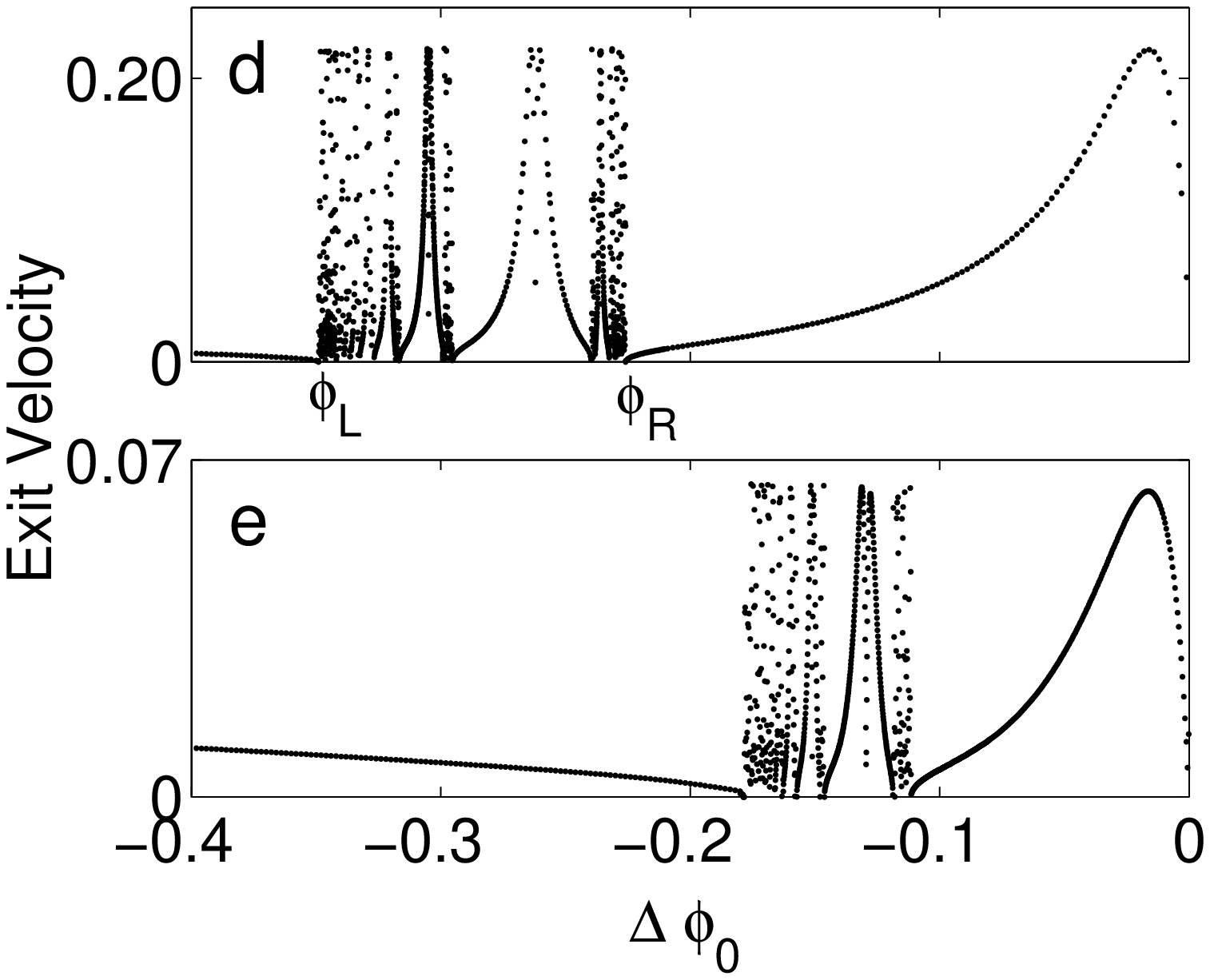}
 \includegraphics[width=40mm,height=31mm]{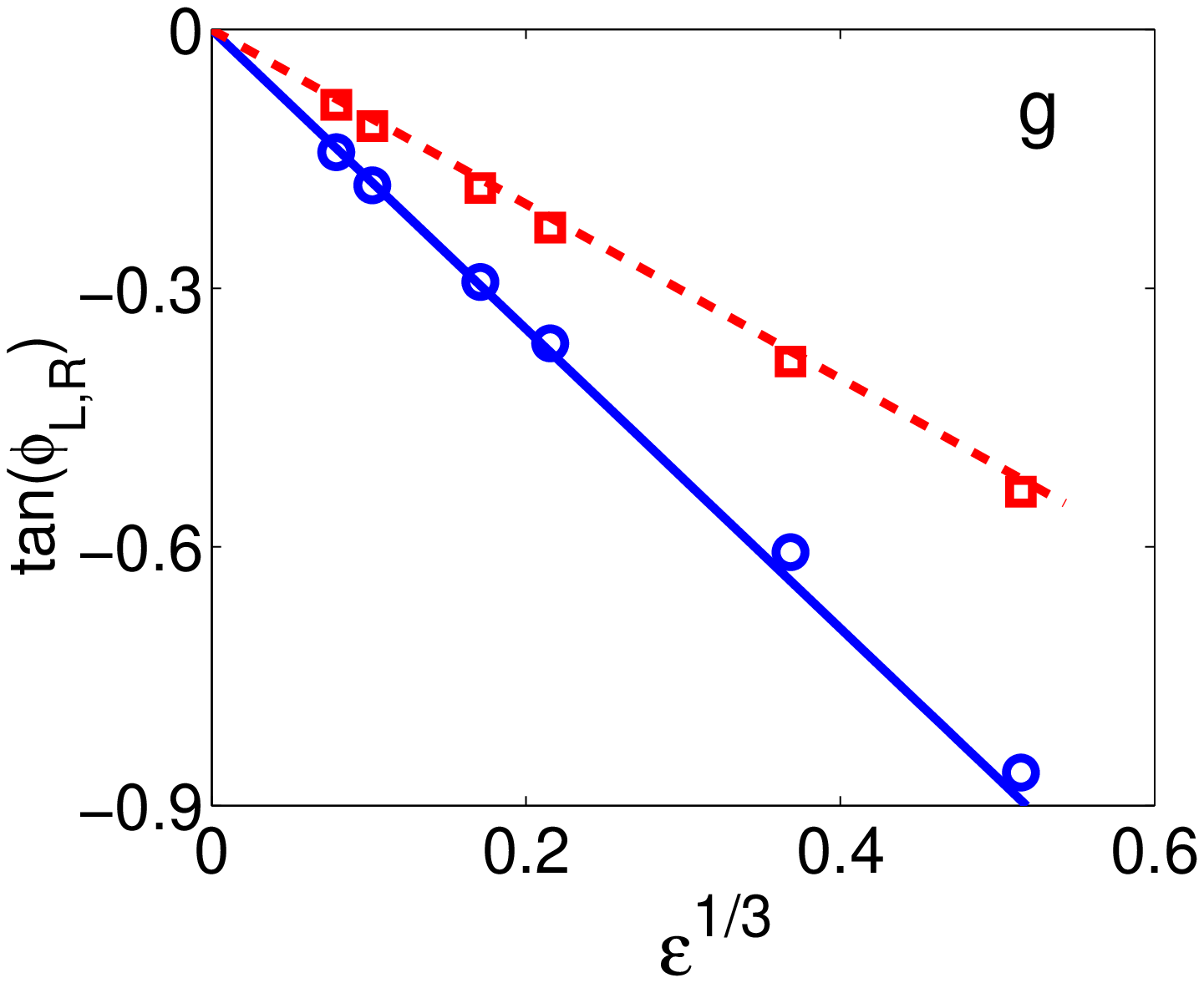}
\caption{(color online) (a,b) The exit-velocity versus
initial-phase-difference graphs from (a) PDE simulations and (b) map
predictions; (c) the map's $|\tilde{M}_\infty|$ graph in the
$(\tilde{Q}_0,\tilde{M}_0)$ plane; (d,e) exit-velocity versus
initial-phase-difference graphs of the PDE for $\varepsilon=0.01$
and $0.001$ respectively; (f) plots of the accumulation points
$\phi_L$ and $\phi_R$ versus $\varepsilon$ from PDE simulations
(circles and squares) and map predictions (solid and dashed lines).
\label{compare}}
\end{figure}
Now we compare this map's predictions with direct PDE simulations.
Here we take the cubic-quintic nonlinearity in (\ref{cqNL}) with
$\gamma$ and initial conditions as for Fig. 1(a). For the map, we
iterate it to infinity to get $\tilde{M}_\infty$ (in practice, 500
iterations performed), which in turn gives $|\dot{\zeta}_\infty|$
from formula (\ref{exit_V}). With the variable scalings
(\ref{scaling1}) and (\ref{scaling2}) considered, the exit velocity
graph predicted from the map (\ref{nmap1})-(\ref{nmap2}) is shown in
Fig. 3(b), while that from the PDE simulations is shown in Fig.
3(a). Comparing these two graphs, it is clear that the map gives a
good replication of the PDE's fractal structure both qualitatively
and quantitatively.

The separatrix map (\ref{nmap1})-(\ref{nmap2}) exhibits a fractal
structure in the graph of $|\tilde{M}_\infty|$ as a function of
initial values $(\tilde{Q_0},\tilde{M_0})$, which is displayed in
Fig. 3(c) (with $\mbox{sgn}(\varepsilon)=1$). This fractal of the
map completely determines the fractal structures in the PDEs
(\ref{GNLS}). For instance, for initial conditions (\ref{ic}), the
corresponding initial values of the map are
\begin{equation}\label{iv}
\tilde{Q}_0=\tilde{M}_0=-2^{-1}(\pi/\varepsilon)^{1/3}\tan(\Delta\phi_0).
\end{equation}
As $\Delta\phi_0$ varies, Eq. (\ref{iv}) gives a parameterized curve
in the $(\tilde{Q}_0, \tilde{M}_0)$ plane, which is the white
straight line in Fig.3 (c). This line cuts cross the fractal of the
map in Fig. 3(c), and the intersection is precisely the fractal
structure as observed in Fig. 1 for the PDE [see also Fig. 3(a,b)].
The scaling laws for fractals of the PDEs can be readily derived
from Eq. (\ref{iv}) and Fig. 3(c). Let $P_1$ and $P_2$ denote the
two accumulation points of the map's fractal, which are
$(\tilde{Q}_0, \tilde{M}_0)=(1.271, 1.271)$ and $(0.741, 0.741)$ as
marked in Fig. 3(c). The corresponding accumulation points $\phi_R$
and $\phi_L$ in the fractal structures of the PDEs are marked in
Fig. 3(d). Here $\phi_L$ and $\phi_R$ are the left and right ends of
the fractal region. Then according to scalings (\ref{iv}), we find
that $\phi_R=-\mbox{atan}(1.482\varepsilon^{1/3}/\pi^{1/3})$, and
$\phi_L=-\mbox{atan}(2.542\varepsilon^{1/3}/\pi^{1/3})$. Hence the
map analytically predicts that the fractal region of the PDE shrinks
to $\Delta \phi_0=0$ as $\varepsilon\to 0$, and its shrinking speed
is proportional to $\varepsilon^{1/3}$ for $\varepsilon \ll 1$. This
is precisely what happens. To illustrate, we choose two $\gamma$
values 0.0029 and 0.0003 in Eq. (\ref{cqNL}), which correspond to
$\varepsilon=0.01$ and $0.001$ respectively. The fractal structures
of the PDEs for these $\gamma$ values are displayed in Fig. 3(d,e).
It is seen that the fractal region indeed shrinks as $\varepsilon$
decreases. We further recorded the $\phi_L$ and $\phi_R$ values in
the PDE fractals at a number of $\varepsilon$ values, and the data
is plotted in Fig. 3(f). The theoretical formulae of $\phi_L$ and
$\phi_R$ above are also plotted for comparison. It is seen that the
PDE values and the map's analytical predictions agree perfectly,
confirming the asymptotic accuracy of the map
(\ref{nmap1})-(\ref{nmap2}).


In summary, we have asymptotically analyzed weak solitary wave
interactions in the generalized nonlinear Schr\"odinger equations
and obtained a simple universal map. This map gives a complete
analytical characterization of universal fractal structures in these
wave interactions. We expect that this work will stimulate research
in other physical systems where weak solitary wave interactions
arise, such as nonlinear optics, water waves and Bose-Einstein
condensates.



\begin{thebibliography}{99}

\bibitem{Ablowitz} M.J. Ablowitz and H. Segur, \emph{Solitons and
the Inverse Scattering Transform}, SIAM, Philadelphia, 1981.

\bibitem{Hasegawa} A. Hasegawa and Y. Kodama, \emph{Solitons in Optical
Communications}, Clarendon, Oxford, 1995.

\bibitem{Kivshar} Y. Kivshar and G. Agrawal, \emph{Optical Solitons}, Academic Press, San Diego, 2003.

\bibitem{Karpman_Solovev} V. I. Karpman and V. V. Solov'ev, Physica D 3, 142 (1981);
K.A. Gorshkov and L.A. Ostrovsky, Physica D 3, 428 (1981).

\bibitem{Campbell} D.K. Campbell, J.S. Schonfeld, and C.A.
Wingate,  Physica D 9, 1 (1983);  M. Peyrard and D.K. Campbell,
Physica D 9, 33 (1983).

\bibitem{anninos}
P. Anninos, S. Oliveira, and R. A. Matzner, Phys. Rev. D 44, 1147
(1991).

\bibitem{Kivshar_Fei} Y. S. Kivshar, Z. Fei, and L.  V\'{a}zquez, Phys. Rev. Lett.
67, 1177 (1991);  Z. Fei, Y. S. Kivshar, and L. V\'{a}quez, Phys.
Rev. A 45, 6019 (1992).

\bibitem{YangTan} J. Yang and Y. Tan, Phys. Rev. Lett. 85, 3624 (2000);
Y. Tan and J. Yang, Phys. Rev. E. 64, 056616 (2001).

\bibitem{Dmitriev} S.V. Dmitriev, Yu.S. Kivshar, and T. Shigenari,
Phys. Rev. E 64, 056613 (2001); S.V. Dmitriev and T. Shigenari,
Chaos 12, 324 (2002).

\bibitem{zhuyang} Y. Zhu and J. Yang, Phys. Rev. E, 75, 036605 (2007).


\bibitem{Goodman_Haberman} R. H. Goodman and R. Haberman,
SIAM J. Appl. Dyn. Sys. 4, 1195 (2005); R. H. Goodman and R.
Haberman, Phys. Rev. Lett. 98, 104103 (2007).


\end{thebibliography}
\end{document}